\documentclass[a4paper,twoside,14pt]{article}

\usepackage[utf8]{inputenc}

\usepackage[english]{babel}
\usepackage{subeqnarray}
\usepackage{url}
\usepackage{lipsum} 
\usepackage[]{mdframed}

\usepackage{graphicx,psfrag}
\usepackage{fixltx2e,amsmath}
\usepackage{xcolor}
\usepackage{amssymb}
\usepackage[final]{pdfpages}
\usepackage{bbold}
\usepackage{dsfont}
\usepackage{pgfplots}
\usepackage{subfig}
\usepackage{algorithm}
\usepackage{algorithmic,placeins}
\usepackage[font=small,labelfont=bf]{caption}

\usepackage[font={footnotesize}]{caption}

\usepackage[utf8]{inputenc}
\usepackage[T1]{fontenc}
\usepackage{xcolor,amsmath,amssymb}
\usepackage{geometry}
\usepackage{graphicx}
\usepackage[sort&compress,numbers]{natbib}
\usepackage{enumerate}
\usepackage{wasysym}
\usepackage[textsize=footnotesize]{todonotes}

\usepackage{hyperref}
\hypersetup{
    colorlinks=true,
    linkcolor=red,
    filecolor=magenta,      
    urlcolor=cyan,
}

\usepackage{pgfplots}

\definecolor{darkblue}{rgb}{0,0,0.6}
\definecolor{darkred}{rgb}{0.6,0,0}

\usepackage{hyperref}

\newcommand{\beq}{\begin{equation}} \newcommand{\eeq}{\end{equation}}

\newcommand\be{\begin{equation}}
\newcommand\bea{\begin{eqnarray} \nonumber }
\newcommand\ee{\end{equation}}
\newcommand\eea{\end{eqnarray}}

\usepackage{authblk}

\usepackage[bitstream-charter]{mathdesign}

\usetikzlibrary{pgfplots.groupplots}

\usepackage{xcolor}

\graphicspath{{figures/}} 
\pgfplotsset{compat=1.18}

\begin{document}

\title{Why is the Dynamics of Glasses Super-Arrhenius?}
\author[1,2]{Jean-Philippe Bouchaud}
		\affil[1]{CFM, 23 rue de l'Universit\'e, F-75007 Paris, France}
  \affil[2]{Acad\'emie des Sciences, Quai de Conti, F-75006 Paris, France}
		
\date{December 2023}
\maketitle

\abstract{The steep increase of the relaxation time of glass forming liquids upon cooling is traditionally ascribed to an impending entropy crisis: since the system has ``nowhere to go'',  dynamics must come to a halt. This classic argument, due to Adam \& Gibbs, has been bolstered and refined by the development of the Random First Order Transition (RFOT) theory, which fares remarkably well at reproducing most salient experimental facts of super-cooled liquids. All static predictions of RFOT, including the existence of a point-to-set length and the role of pinning sites, have been vindicated by detailed numerical simulations. Yet, there is no consensus that the basic mechanism explaining the glass transition is the one captured by RFOT. Strong doubts have emerged following the observation that adding or removing kinetic constraints can change the relaxation time by orders of magnitude, while leaving thermodynamics unchanged. This is at odds with the idea of a one-to-one mapping between excess entropy and relaxation time. 

In the following discussion paper presented at the Solvay conference in October 2023, we review areas of consensus and dissent of RFOT with other competing theoretical proposals, and propose possible paths for (partial) reconciliation. We further argue that extensive numerical simulations of the non-linear susceptibility of glasses, in particular in the aging regime, should shed important light on the mechanism at the origin of the super-Arrhenius behaviour of the relaxation time. In any case, more imagination is still needed to come up with experimental, theoretical or numerical ideas that
would allow to finally settle the question of why glasses do not flow.} 

\section{The Glass Conundrum}

As is well known, the {\it glass conundrum} is that supercooled liquids slow down at a fantastic pace when temperature is reduced, without any apparent change in their structural characteristics, i.e. the way molecules organize in space. In the case of Ortho-Terphenyl, for example, the relaxation time increases by a factor $10^{10}$ as temperature drops by a mere $10\%$!

A host of empirical facts, accumulated since the 1960's, suggest that the slowdown is intimately related to the loss of ``excess'' or ``configurational'' entropy per molecule $S_{\text{xs}}$ (i.e. the non-vibrational part of entropy counting the number of possible arrangements of molecules that are in mechanical equilibrium). More precisely, the logarithm of the relaxation time $\tau$ appears to be proportional to $S_{\text{xs}}^{-1}$, with $S_{\text{xs}}$ dropping precipitously when temperature approaches the Kauzmann temperature $T_K$ \cite{Nagel_Angell}. This results in a fast, super-Arrhenius growth of $\tau$ that appears to diverge at $T_K$, characteristic of ``fragile'' glasses \cite{Angell}. One observes that the stronger the drop in specific heat at the glass transition $T_g$ (where the system is no longer in equilibrium over experimental time scales), the more fragile is the glass, and the more non-exponential in time is the relaxation process \cite{Xia}. 

Such correlations between thermodynamics and dynamics have been for many years a guiding principle for theorists. A famous early argument by Adam \& Gibbs suggests that the cooperative motion of an increasingly larger and larger number of molecules is needed to relax the liquid as it gets colder, leading to increased effective energy barriers and therefore super-Arrhenius behaviour \cite{AG}. In a nutshell, Adam \& Gibbs reason that for the system to evolve, at least one other configuration, different from the current one, must be available. Hence, a number $n$ of molecules must be involved, such that $n S_{\text{xs}} \geq \log 2$ for at least two configurations to be available. If one then postulates that the associated energy barrier is proportional to a power of $n$, one immediately concludes that a) the super-Arrhenius behaviour is indeed {\it caused} by the drop of entropy, and b) the growth of the energy barrier cannot occur without the growth of a cooperative volume. 

Although inspiring, the Adam-Gibbs argument is, of course, full of loopholes. But its physical content matches, with appropriate modifications, the conclusions of the Random First Order Transition (RFOT) theory proposed 25 years later by Kirkpatrick, Thirumalai and Wolynes \cite{KTW1,KTW2,KTW3}. Although initially based on a highly stylized mean-field spin-glass model, most qualitative (and sometimes quite subtle) {\it static} predictions of the RFOT theory have been confirmed by more realistic theoretical models, such as lattice glass models \cite{biroli_mezard} or hard spheres in high dimensions \cite{zamponi_book}, and by detailed numerical simulations (see \cite{CRAS} and G. Biroli and F. Zamponi, this volume). In particular, the existence of a cooperative length scale $\ell$  {\it à la} Adam-Gibbs is now firmly established, together with its precise interpretation as a static {\it point-to-set} correlation length \cite{Montanari-Semerjian}, that grows as $S_{\text{xs}}^{-\text{power}}$ \cite{Bou04, RFOT1}. Remarkably, a phenomenological {\it dynamical} extension of RFOT to describe supercooled liquids allows one to successfully account for most empirical results \cite{KTW3,RFOT1,wolynes-review,RFOT2}, and is supported by several numerical studies (see e.g. \cite{Cam09,karmakar,pts1}).  

This could have been the end of the story, with more and more people -- after digesting the many subtleties of the theory -- agreeing that RFOT provides the right canvas to think about the glass transition. But this is not what happened: not only did theorists come up with several plausible alternative scenarii that cannot be rejected out of hand \cite{KCM,bbreview,Tarjus}, but also deeply troubling facts emerged from state-of-the-art numerical simulations, which are not naturally explained by the RFOT theory -- see below. It is fair to say that at this stage, there is no consensus on the basic mechanism responsible for the glass transition. 

At one extreme of the theoretical spectrum lies the RFOT scenario \cite{wolynes-review,RFOT2}: the slowdown of glasses is a consequence of an incipient phase transition towards a state characterized by some ``amorphous long-range order''. This sounds like an oxymoron, but in fact accurately describes the physics of spin-glasses. Much as in glasses, instantaneous snapshots of the spin configurations seem featureless both above and below the phase transition. But whereas there is no long range transmission of information at high temperature, the spin-glass phase is \textit{rigid}, {{as is the glass phase}} \cite{yoshino}, in the sense that localised perturbations have a long range effect on the system \cite{anderson}. 

At the other extreme of the spectrum, one finds ``local'' theories that ascribe the slowdown to the growth of purely local barriers that impede elementary moves, without having to invoke cooperative motion as a mechanism for barrier growth \cite{shoving, WC, Wnew}. The ``shoving model”, for
instance, proposes that the chief physical ingredient driving the glass transition is the growth of the plateau shear modulus, $G_{\text{hf}}$, which makes even local moves progressively more difficult -- see \cite{dyre_new} for a very nice recent review. The growth of the activation barrier to flow would then simply mirror that of $G_{\text{hf}}$ \cite{shoving}, without involving any growing cooperative volume \cite{WC}.

Intermediate pictures are also on the market, where thermodynamics play no role but so-called ``kinetic constraints'' require collective motion for relaxation to take place \cite{Ritort_Sollich, KCM, Keys}. In these theories, the progressive logjam of super-cooled liquids is due to a rarefaction of local ``mobility defects” that act as facilitators for structural rearrangements. In this scenario thermodynamics only plays a minor role, or even no role at all. The glass is but a liquid that cannot flow because of kinetic constraints, but there is no driving force towards any kind of locally preferred structure or amorphous order.

In the following discussion, we will briefly review the challenges faced by the RFOT theory and the areas of both consensus and dissent between the competing pictures of the glass transition. At the heart of the debate lies the timeworn distinction  between {\it correlation} and {\it causation}. Whereas the existence of Adam-Gibbs correlations between thermodynamics and dynamics is beyond any reasonable doubt, the issue is whether these correlations are accidental and can be explained by some anecdotal consequence of the ultimate cause of glassy slowing down \cite{Wyart}, or if these correlations indeed reveal such ultimate cause and can be brandished as trophies of RFOT. At a deeper level, one is faced with the  question of how to validate unambiguously one particular theory of the glass transition and eliminate the others (or perhaps unify some of them within a common framework). Even if lots of interesting -- and sometimes groundbreaking -- work has been done in the last decades, it looks as if more imagination is still needed to come up with experimental, theoretical or numerical ideas that would break the deadlock.  

\section{RFOT: Successes and Challenges}
\label{sec:RFOT}
Over the last two decades, several aspects of the RFOT theory have received confirmation both from analytical calculations on solvable cases (see e.g. \cite{biroli_mezard,franz,zamponi_book,Tarjus_R,yoshino}) and from atomistic simulations. Of particular importance is the confirmation that metastable states with extensive configurational entropy $S_{\text{xs}}$ play an important role, as revealed by the behaviour of the Franz-Parisi potential \cite{FP-potential}, and its strong {\it local} correlation with the relaxation time \cite{berthiernew}, which suggest that the Adam-Gibbs argument holds even locally. Microscopic calculations have furthermore revealed a deep link between RFOT and the Random Field Ising Model \cite{wolynes-rfim,franz-rfim,RFIM1,RFIM2}, with quite non-trivial predictions about the critical behaviour of the Franz-Parisi potential that are in surprisingly good agreement with numerical simulations \cite{guiselin}. 

Crucially, the existence of a non-trivial point-to-set length $\ell$ and its growth when $S_{\text{xs}}(T)$ decreases, which are arguably the most important RFOT predictions, are now firmly established by numerical simulations \cite{Cavagna,dzero,franz,pts1,scalliet}. However, the role played by such a growing static length scale in the dramatic slowing down of super-cooled liquids has recently been the subject of renewed qualms \cite{WC, Wnew}. 

Let us first recall the argument relating $\ell$ to the relaxation time of the liquid. Consider the situation of particles confined in a cavity with frozen amorphous boundary conditions. When the cavity
radius $R$ is less than $\ell$, the liquid inside the cavity is frozen too, in the sense that only a small subset of configurations has a significant weight in the Boltzmann measure. When
$R > \ell$, on the other hand, the number of metastable configurations
becomes so large that even when most of them
are incongruous with the boundary conditions, the cavity is driven by entropy into the liquid state. In other words, relaxation of the density field cannot occur unless the radius of the cavity is of the order of, or larger than $\ell$. Note that this statement is independent of the actual dynamics driving the system (provided of course it obeys detailed balance). 

Within the RFOT scenario, the free-energy barrier $B$ for rearrangements in such a cavity of size $\ell$ is argued to grow as
\begin{equation} \label{eq:Barrier}
   B(T) \sim \Delta(T)\, \ell^\psi 
\end{equation}
where $\Delta(T)$ is a temperature dependent energy scale and $\psi$ is a certain exponent. Associating the relaxation time $\tau$ with $\exp(B/T)$, one naturally accounts for both (a) the empirically observed Adam-Gibbs correlation between configurational entropy and relaxation time; and (b) the strongly non-Arrhenius, Vogel-Fulcher-type increase of $\tau$ in fragile liquids. 

However, the RFOT scenario for the dynamics of glasses is jeopardized by at least three observations:
\begin{enumerate}
    \item A large part of the growth of the effective energy barrier $B(T)$ when $T$ decreases is accounted for by the stiffening of local ``cages'', i.e. the growth of the high frequency shear modulus $G_{\text{hf}}(T)$ when $T$ is decreased. In other words, the curvature of $\log \tau$ when plotted as a function of $1/T$, characteristic of fragile liquids, is much reduced (but still clearly present) when plotted as a function of $G_{\text{hf}}/T$ \cite{Torchinsky}. This is essentially the content of Dyre's ``shoving'' model \cite{shoving, dyre_new}: in order to move, molecules have to shove away their neighbours, with an elastic cost $\propto G_{\text{hf}}$. But this means that the role of the point-to-set length $\ell$ in the growth of $B(T)$ (see Eq. \eqref{eq:Barrier}) would be minor, or even irrelevant, compared to that of the energy scale $\Delta \propto G_{\text{hf}}$. 
    \item More importantly, the effective energy barrier $B$ appears to be extremely sensitive to the addition or removal of kinetic constraints. For example, one knows that local swaps of particles of different radii can decrease the relaxation time by orders of magnitude \cite{swap}, whereas restricting the direction of motion of molecules can increase the relaxation time by orders of magnitude \cite{Gava}. But if this is the case, how can {\it thermodynamics} be so relevant in determining the relaxation time \cite{WC}, as postulated by the Adam-Gibbs/RFOT theory?\footnote{An important assumption here is that the liquids for which the swap algorithm works are representative of real glass-forming liquids. Although there are no indication so far that this is not the case, one should probably be cautious about drawing too strong conclusions.} 
    \item Wyart \& Cates \cite{WC} further argue that the Stokes-Einstein decoupling between self-diffusion and collective relaxation should be much stronger than experimentally observed if local kinetic constraints were not the dominant effect.\footnote{The self-diffusion constant is $\sim 10^3$ larger than inferred from the value of $\tau$ at $T=T_g$. If particles where {\it individually free to move} but collectively trapped, this enhancement factor should be closer to $10^{15}$ \cite{WC}. For counter-arguments, see \cite{crumbling}.}
\end{enumerate}

Such arguments need to be seriously addressed by the RFOT team (in which the author counts himself). In particular, the RFOT theory makes universal predictions for $\Delta$ and $\psi$ in Eq. \eqref{eq:Barrier} \cite{KTW3,RFOT1} that are subject to debate and do not naturally account for the strong dependence of $\tau$ on kinetic constraints \cite{swap,Gava}. 

But in fact, as discussed in the next section, there are already several areas of consensus that are worth emphasizing, as they offer some clues about a possible -- at least partial -- reconciliation of different theories and suggest directions worth exploring further. 

\section{Areas of Partial Consensus}

\subsection{The Mode-Coupling Temperature}

\begin{figure}[!ht]
\centering
\includegraphics[width = 9cm]{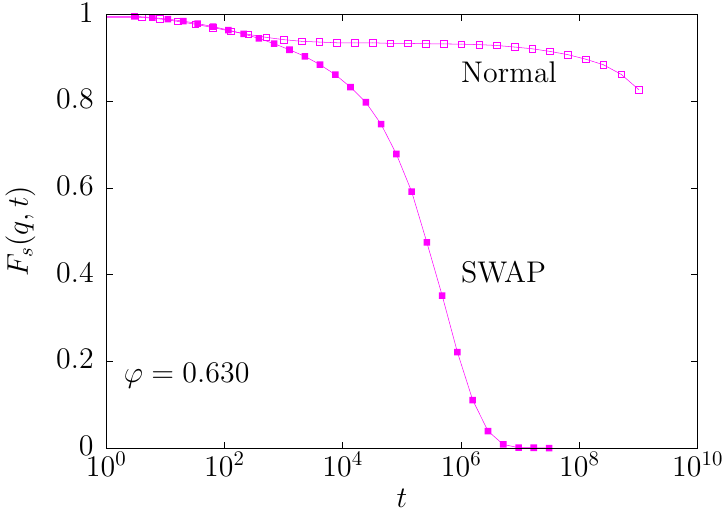}
\caption{Self-intermediate scattering function for a three dimensional polydisperse hard-sphere system of volume fraction $\varphi$, with or without SWAP dynamics. This clearly illustrates the idea of ``crumbling'' metastability: the plateau corresponding to local rigidity -- that extends over more than $4$ decades in time for the ``normal'' dynamics -- completely disappears in the presence of swaps. From \cite{crumbling}.}
\label{figcorr}
\end{figure}

Let us first note that all viable theories derive or postulate the existence of a ``Mode-Coupling'' temperature $T^\star$ below which \textit{metastable states} appear, corresponding to ``cage formation'' inside which molecules become self-consistently trapped, at least temporarily, and only escape through thermal activation -- as postulated long ago by Goldstein \cite{goldstein}. Indeed, one can only speak about activation barriers and slow dynamics if the system is locally stable, i.e. if some local {\it rigidity} sets in \cite{yoshino} and prevents free flow.\footnote{See also the discussion in \cite{coslovich}.} This corresponds to the onset of a characteristic {\it plateau} regime in the structural relaxation, as shown in Fig. \ref{figcorr}.  

RFOT theory predicts that below $T^\star$, some amorphous order sets in and extends over the point-to-set length $\ell$, accompanied by the appearance of a high-frequency shear modulus $G_{\text{hf}}(\omega \sim \tau^{-1})$ that grows sharply (as $\sqrt{T^\star - T}$ in mean-field) as temperature is reduced. 

``Local'' theories claim that such a rapid elastic stiffening of the cages is enough to explain the non-Arrhenius behaviour of the relaxation time $\tau$, whereas RFOT argues that relaxation has to involve $n \sim \ell^3 $ molecules, leading to a collective energy barrier that grows as in Eq. \eqref{eq:Barrier}. 

Kinetically Constrained Models (KCM, \cite{Ritort_Sollich}) lie mid-way: local rigidity is taken for granted as it tacitly justifies why dynamics is shackled (``kinetically constrained''). Rare mobile regions cannot move freely and must conspire to propagate in space. This becomes more and more difficult as temperature is reduced, which in turn leads to non-Arrhenius behaviour below an onset temperature where kinetic constraints become dominant \cite{KCM, Keys}, again identified as a Mode-Coupling temperature. For KCM advocates, the reduction of configurational entropy is not the core mechanism for dynamical arrest and Adam-Gibbs correlations are spurious, as for local theories.\footnote{On this point, see the discussion in \cite{BBT}.} But in line with RFOT, non-Arrhenius KCM in fact rely on some degree of dynamical cooperativity, as we discuss in section \ref{sec:recap} below.    

\subsection{Is the Efficiency of SWAP Incompatible with RFOT?}
\label{sec:efficiency}
As was mentioned above, the extreme sensitivity to the relaxation time $\tau$ on (local) dynamical rules suggests that entropic considerations cannot explain the slowing down supercooled liquids. In this section, we explain how kinetic constraints can have a dramatic effect on the Mode-Coupling $T^\star$, and therefore on the effective energy barrier $B$, in a way fully compatible with RFOT.    

In mean-field RFOT, metastable states can be defined unambiguously, independently of the dynamics, because energy barriers separating them are infinite. In real systems, however, the fact that a set of micro-states forms a metastable state depends both on the dynamical rules and on a timescale. This timescale should be long enough to allow equilibration among such a set of micro-states and yet be short enough for not allowing the system to escape from that set. Importantly, such a separation of timescales may hold for one set of dynamical rules and not for another, for example dynamics with or without swaps of particles of different radii.  

A minimum requirement for local metastability is that the Hessian of the configuration energy computed inside a cavity of size $\ell$ should be definite positive. Now, the SWAP algorithm effectively allows the radius of the particles to fluctuate \cite{brito}, thereby increasing the number of degrees of freedom and the dimension of the Hessian matrix. Some unstable directions can therefore appear, that would not exist without swaps. Hence, states that are metastable without swaps may lose their local rigidity when swaps are allowed. A signature of this ``crumbling metastability'' \cite{crumbling} can be seen in Fig. \ref{figcorr}: the two-step relaxation curve, signalling the formation of local cages, is completely wiped out by swaps. There is no longer any barrier preventing motion -- this is actually precisely why the SWAP algorithm is so effective!  

The excess of unstable directions means that the appearance of local metastability is pushed to lower temperatures, i.e. $T^\star_{\text{swap}} < T^\star$. Even though the point-to-set length $\ell$ is independent of the dynamics and therefore still exists, in principle, between $T^\star_{\text{swap}}$ and $T^\star$, there is no collective activation barrier in this regime, i.e. $\Delta(T)=0$ (see Eq. \eqref{eq:Barrier}). The above scenario, that explains the success of the SWAP algorithm in terms of a downward shift of the Mode-Coupling temperature, has been proposed in different incarnations in Refs. \cite{ikeda,brito,szamel,crumbling}.  

Below $T^\star_{\text{swap}}$ and for large enough $\ell$, one expects the (free-) energy barriers given by Eq. \eqref{eq:Barrier} to be independent of the dynamics. Hence the relaxation time $\tau_{\text{swap}}(T)$ should in fact massively increase below $T^\star_{\text{swap}}$ to catch up the no-swap value $\tau(T)$. A consequence of this picture is that swap dynamics should lead to anomalously fragile behaviour at low temperature. Hints of such an increased fragility can be seen in Ref. \cite{crumbling}, Fig. 2, bottom graph.
Conversely, adding extra constraints on the dynamics, as proposed in \cite{Gava}, should reduce the fragility of the system. It would be interesting to test these predictions more extensively.   

\subsection{Cooperativity vs. Facilitation}
\label{sec:CoopvsFaci}

As discussed in the introduction, the idea of cooperative motion idea dates back to Adam \& Gibbs and has been made precise, within RFOT, by the concept of point-to-set (PTS) static correlations: the PTS length $\ell$ sets the scale of the smallest cavity such that the molecules inside the cavity are no longer ``pinned'' by the boundary conditions outside the cavity. In other words, below $T^\star$ the liquid is made of small ``glassites'' of size $\ell$ that must collectively rearrange for relaxation to occur. One area of dispute is whether the time to do so is dominated by the collective energy barriers  $B(\ell)$ (as in Eq. \eqref{eq:Barrier}) or by local energy barriers, for example the elastic energy $\propto G_{\text{hf}}$ needed to ``shove away'' nearby molecules.

A different concept is that of ``facilitation'', which captures the fact that a relaxation event taking place around point $\vec r$ in space can trigger another relaxation event somewhere else through, e.g., elastic coupling as envisaged originally by Hébraud \& Lequeux in their seminal paper on elasto-plastic dynamics \cite{HL}. Such a mechanism allows the whole system to relax much faster, since locally slow regions can be unlocked by a locally fast region rearranging nearby. It also leads to {\it dynamical correlations}, as measured by 4-point correlation functions \cite{DH1}, that extend over length scales $\xi \geq \ell$. 

That such a facilitation mechanism exists in both model and real glass-formers is beyond any doubt, see for example \cite{candelier,Keys,chacko}, and in particular Ref. \cite{scalliet_guiselin} which explores low temperatures and long-time scales, and Ref. \cite{Tahaei} in the context of elasto-plastic models. Interestingly, facilitation explains the ubiquitous ``excess (high frequency) wing'' in the dissipative part of the linear susceptibility $\chi''(\omega)$ \cite{excess_wings,scalliet_guiselin}. To wit, the high-frequency wing is the footprint of a broad (power-law tail) distribution of local relaxation times, which gets truncated at low frequencies as the fastest regions unlock the very slow ones.  

Facilitation effects have only been tangentially discussed within RFOT, although it has always been clear than a relaxing glassite would change the boundary conditions of its neighbours and induce a local propagation of relaxation, speeding up the whole system (see e.g. \cite{Xia,WW} and the explicit discussion in \cite{RFOT2}, section 2.4.3). Such mechanism is in fact at the heart of the ``crumbling metastability'' argument of \cite{crumbling} and probably plays a major role in the success of SWAP -- see \cite{berthier2023} for a recent discussion. However, longer range, elasticity mediated facilitation has not been considered in RFOT papers and seems to play an important role, see \cite{Tahaei}.

Let us finally recall that the Mode-Coupling theory (MCT) makes some predictions concerning the dynamical correlation length $\xi$ \cite{IMCT,DH2}, which is found to diverge when $T \downarrow T^\star$, indicating that the relaxation process becomes more and more correlated (but not cooperative) as temperature is decreased above $T^\star$. However, RFOT now predicts {\it three} different length scales when $T < T^\star$: one, implied by MCT, decreases when $T \downarrow$ but now only describes the $\beta-$relaxation, i.e. the pre-plateau regime; the second one is the point-to-set length $\ell$ which governs the cooperative relaxation process, and finally the dynamical correlation length $\xi$ that reflects the facilitation process, which is unrelated to the Mode-Coupling mechanism in this regime.

\subsection{Recap: Scenarii for Glassy Slowdown} 
\label{sec:recap}

Summarizing what we have discussed so far, {\it all} theories \footnote{Including the frustration based theory of Tarjus et al. \cite{Tarjus_R} based on the idea of an avoided phase transition.} assume that glassy dynamics is the result of a rather sudden ``cage formation'' process around a temperature $T^\star$, above which molecules freely flow as in a liquid, and below which single particle motion is hindered. Relaxation is only possible through rare  elementary activated events (``activons'') that either trigger avalanches of other rearrangements, or slowly propagate through space and progressively unlock the regions they visit. 

For both local and RFOT theories, the time scale associated to activons grows faster than Arrhenius, in the former case because {\it local} energy barriers grow as temperature decreases, but without any notion of cooperativity; in the latter case because the number $n \propto \ell^d$ of molecules required to move in sync grows as temperature decreases.\footnote{Here and elsewhere, $d$ is the space dimension.} In other words, activons are point-like in the local picture and extended over a growing length scale $\ell$ in the RFOT picture. In the KCM scenario, the energy $J$ needed to create an elementary mobility defect is taken to be independent of temperature, so the equilibrium density of such defects is $\rho \propto e^{-J/T}$. 

In local/RFOT scenarii, all the ``heavy lifting'' is done at the activon level; facilitation, if anything, speeds up relaxation elsewhere by prematurely unlocking the slowest regions. In the KCM scenario, on the other hand, it is the anomalously slow propagation of mobility defects that eventually leads to super-Arrhenius time scales. In this sense, ``facilitation'' is a bit of a misnomer in the case of KCM. 

More precisely, let us consider a model where the distance travelled by a mobility defect grows like $r \sim t^{1/z}$, where $z$ is a temperature dependent dynamical exponent. When $z > d$, exploration of space is compact. Therefore, all regions have had a chance to decorrelate when $\rho r^d \sim 1$, which translates into a relaxation time $\tau_{\text{kcm}}$ given by \cite{Ritort_Sollich, Keys}
\begin{equation}
    \tau_{\text{kcm}} \sim e^{\frac{Jz(T)}{dT}}.
\end{equation}
In some models, like the ``East'' model \cite{Ritort_Sollich}, the temperature dependence of the dynamical exponent is given by $z(T)=T_0/T$, which leads to slower and slower expansion of the mobile regions and in turn a super-Arrhenius growth of the relaxation time, as $\log \tau \propto T^{-2}$ \cite{Keys}. Glassy slowing down in KCM is thus the result of two effects: the rarefaction of mobility defects, and their more and more sluggish progression in space. 

Note that although there is no explicit mention of ``amorphous order'' in the KCM picture, the assumption that particles cannot move unless a mobility defect passes by is in fact similar, physically, to the RFOT idea that the dynamics on scales $< \ell$ is frozen. In a sense, the KCM picture might be reinterpreted as postulating that the activation path needed to unlock a glassite of size $\ell$ is through the anomalous diffusion of point-like defects, with an effective energy barrier $B(\ell) \sim \log \ell$, see Ref. \cite{nishikawa} for a explicit illustration in the context of lattice glasses. 

So the short summary of the current discussion about the physics of glassy systems is: 
\begin{mdframed}
    {\it Is the glassy relaxation time dominated by “activons” or the slow diffusion of mobility, or by a mixture of both? Are activons ultra-local or extended over a growing length scale $\ell$?}
\end{mdframed}

Before attempting a tentative answer to these questions in the conclusion section, we first want to briefly review the information that one can extract from recent non-linear susceptibility experiments.

\section{Non-linear Susceptibility: A Smoking Gun?}

\begin{figure}[!ht]
\centering
\includegraphics[width = 9cm]{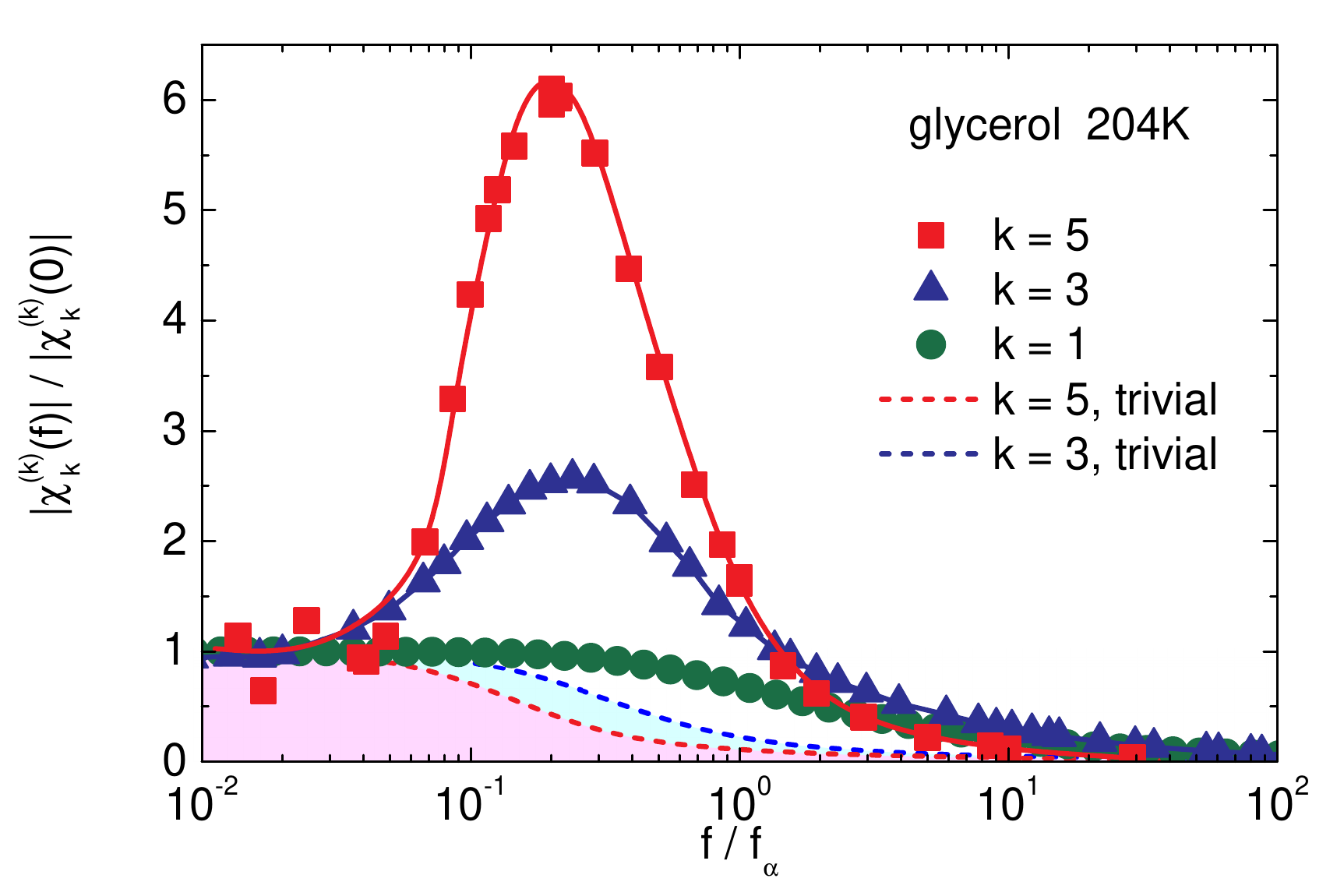}
\caption{Adapted from Ref. \cite{Alb16}. Comparison of the susceptibilities of various orders $\chi_k$ (scaled by their value at zero frequency) in glycerol at $T=204$K $\simeq T_g + 16$K. Two points are noteworthy: as predicted by theory {(i)} the humped shape in frequency is only present for nonlinear susceptibilities $k \geq 3$; {(ii)} the peak is much more pronounced for $\chi_5 \propto \ell^6$ than for $\chi_3 \propto \ell^3$. For comparison the dashed lines show the corresponding featureless curves for the case of an ideal gas of dipoles (see \cite{Alb16} for more details).}
\label{fig3}
\end{figure}

As reviewed in section \ref{sec:CoopvsFaci}, the existence of a point-to-set length, and its growth when temperature is reduced, are well established both theoretically and numerically \cite{Cavagna,dzero,franz,pts1,scalliet}. 
A direct experimental confirmation that super-cooled liquids can become glasses by confining them in small enough frozen cavities formed by the same liquid and/or when pinning a high enough fraction of particles would be a vindication of the basic premises of RFOT (but not necessarily inconsistent with other theories). A direct measurement of $\ell$ in molecular glass formers is obviously extremely difficult, see e.g. \cite{pinning-expt,Das0,Das}. 

An indirect method, based on the idea that frozen glassites should respond \textit{collectively} to an oscillating field, exploits the behaviour of non-linear dielectric susceptibilities \cite{Bou05}. If  glassites are compact, theory predicts \cite{Alb16} that the k$^{\text {th}}$-order dielectric susceptibility $\chi_k$ should peak at a value proportional to\footnote{Note that the linear dielectric susceptibility ($k=1$) is therefore expected not to show any anomalous increase, as in spin-glasses and in agreement with experiments, see Fig. \ref{fig3}.}  $\ell^{3(k-1)/2}$, i.e. $\ell^3$ for the third-order non-linear susceptibility and $\ell^6$ for the fifth-order susceptibility, both of which having between measured by two experimental groups \cite{Alb16}. The peak is expected to be located at a frequency $\omega \sim \tau^{-1}$, since at higher frequencies only fast clusters can follow the field and at lower frequencies glassites have relaxed and the collective response of frozen dipoles is lost. In other words, amorphous order is only transient and non-linear susceptibilities at zero frequencies are expected to behave as in normal liquids. 

These predictions agree quantitatively with experiments, see Fig. \ref{fig3} and  \cite{Alb16}, which can be seen as the best indirect experimental evidence to date of the growth of a static length scale in super-cooled liquids close to the glass transition. In particular, the peak amplitudes of $\chi_{3,5}$ grow as temperature decreases, reflecting the corresponding growth of $\ell$, in a way compatible with Eq. \eqref{eq:Barrier}. 

Hence, these experimental results strongly suggest that molecules in volumes of size $\ell^3$ collectively unlock after a time $\sim \tau$. Purely kinetic theories of the glass transition, where thermodynamics is trivial or plays no role, cannot explain such anomalous non-linear effects -- see \cite{BBL, LadieuNew} for a detailed discussion of this point. But as we noted in the previous section, what would allow an effective description in terms of kinetic constraints only is the fact that the system exhibits some form of non-local rigidity, precisely as predicted by RFOT below $T^\star$, offering a possible reconciliation here.  

Local theories do not deny the possible existence of such local rigidity and the existence of a point-to-set length $\ell$ -- in fact, the anomalous growth of local energy barriers is also ultimately related to the existence of a Mode-Coupling like transition temperature $T^\star$. But the argument is rather that the main contribution to the increase of the effective energy barrier $B$ is local. In other words, as soon as a purely local activated event has taken place, dynamics on scale $\ell$ (which is needed to relax the system) is fast, presumably mediated by facilitation. The collective energy barrier  $\Delta \ell^\psi$ (as in Eq. \eqref{eq:Barrier}) is then somehow by-passed, or subdominant. 

This is of course a possibility. But there is a set of experimental results that in our opinion has not been appreciated enough: the aging behaviour of $\chi_3$ that unambiguously reveals the growth of $\ell$ with the age of the system, during which $\tau$ itself also increases \cite{Bru12}. It turns out that the parametric relation between $B := T \log \tau$ and $\ell$ as the age increases is compatible with Eq. \eqref{eq:Barrier}, with reasonable values of $\Delta$ and $\psi$ \cite{Bru12}. This could again be a fortuitous coincidence; a way to test this would be to measure simultaneously the growth of local energy barriers or of $G_\text{hf}$ with age. We would gain a lot of insights from reliable numerical simulations of $\chi_3$ for realistic glass formers, both in equilibrium and during aging. This would, for a start, allow one to ascertain the amorphous clusters interpretation of the growth of $\chi_3$ and its relation with the point-to-set length (see \cite{LadieuNew} for a recent discussion).

\section{Freewheeling Discussion \& Suggestions}

Let us emphasize again that, quite remarkably, many rather non-trivial predictions of RFOT theory about the \textit{statics} of super-cooled liquids, in particular the existence of an entropy driven point-to-set length $\ell$, have been confirmed by analytical calculations or numerical simulations in the last decade, see \cite{CRAS} and section \ref{sec:RFOT}.

It would be surprising, but of course not impossible, that the presence of extended frozen clusters (``glassites'') has no impact on the dynamics of super-cooled liquids. For one thing, the ultimate justification of using models that abstract away from thermodynamic forces and entirely focus on kinetic constraints is precisely the existence of medium range amorphous order. A way to interpret KCMs would then be that they represent a somewhat ad-hoc effective model for the residual dynamics of a system for which rigidity has emerged at low temperatures/high density. 

Similarly, the reason why local barriers might grow abruptly when temperature is reduced is hard to understand without evoking some kind of collective rigidity transition taking place -- which is in fact exactly what  MCT/RFOT predicts.

The outstanding question is then whether the non-Arrhenius growth of the relaxation time $\tau$ is (a) primarily due to the collective motion of an ever growing number of molecules $n \sim \ell^d$ (RFOT) \cite{KTW3,Bou04,RFOT2}; (b) primarily due to the anomalously slow motion of rare, thermally activated mobility defects (KCM) \cite{KCM,Keys}; or  (c) primarily due to local energy barriers involving the rearrangement a fixed (small) number of molecules (Local) \cite{shoving,WC,Wnew,dyre_new}. 

Of course, these three elements could be partially compatible and mix in various proportion (see e.g. \cite{wolynes_2013}). But note that in scenarii (a) and (c), facilitation actually speeds up the overall relaxation of the system, whereas in scenario (b) anomalously slow diffusion of mobility defects is the mechanism leading to super-Arrhenius relaxation times. 

The following discussion is highly conjectural, and only reflects the author's understanding at the time of writing.
\begin{enumerate}
    \item A possible reconciliation of (a) and (b) is that KCM offers an explicit construction of the activated path leading to decorrelation on scale $\ell$, through the anomalous wandering of point-like mobility defects in an essentially frozen environment. In fact, the surprisingly small value of the exponent $\psi$ (see Eq. \eqref{eq:Barrier}) reported numerically \cite{Cam09,karmakar} and experimentally \cite{ozawa} might suggest such an interpretation -- see also \cite{nishikawa}. It would also provide an alternative explanation for why adding or removing explicit kinetic constraints has such a strong effect on the energy barrier.    
    \item A possible reconciliation of (a) and (c) could be through facilitation: since any local activated rearrangement might trigger other activated events elsewhere, it may well be that relaxation on scale $\ell$, needed for allowing the system to flow, is catalyzed by local events, leading to a much weaker-than-expected dependence of the barrier $B$ on $\ell$. An alternative mechanism is the following: facilitation means that global relaxation only requires the exceptionally fast ``glassites'' to evolve, possibly leading to a much milder dependence of $\log \tau$ on $\ell$.   
\end{enumerate}

However, if the growth of {\it local} barriers is the dominant effect, as in c), one still needs to explain why Adam-Gibbs-like correlations are ubiquitous across so many fragile glass-formers, even if such correlations turn out to be somehow accidental and not causal (see \cite{Wyart} for a possible mechanism). These correlations in fact extend to the relation between of $\ell$ (and not only $S_{\text{xs}}$) and of $\log \tau$, both numerically \cite{Cam09,pts1,berthiernew,guiselin_overlaps} and experimentally \cite{ozawa, Bru12}. Of particular interest in this respect are the non-linear aging experiments of \cite{Bru12} that suggest an even stronger form of {\it out-of-equilibrium} correlation between $\log \tau$ and $\ell$, which both increase with age. The question then is: is relaxation time shorter in the aging regime because $\ell$ is smaller or is it because local barriers are lower? Investigating numerically the joint evolution of non-linear susceptibility, relaxation time and local energy barriers in an aging regime would bring some precious information on this issue. Intuitively, one would expect local quantities (like $G_{\text{hf}}$) to reach their equilibrium value rather quickly, meaning that aging should only manifest itself through the growth of locally ordered clusters, possibly allowing one to disentangle local effects from collective effects. Heating and cooling protocols, along the lines of \cite{cooling}, could also shed light on the core mechanisms at play.

Finally, we wish to suggest three possibly interesting directions that would enhance further our understanding of the glass transition:
\begin{itemize}
\item One is to investigate the fragility of model glasses when one adds or removes kinetic constraints, like in \cite{swap,Gava}. If the RFOT theory is correct, one expects that the shift in the effective MCT transition temperature $T^\star$, as discussed in section \ref{sec:efficiency}, should make SWAP dynamics asymptotically {\it more fragile} and more constrained dynamics {\it less fragile}. Hints of increased fragility under SWAP have been reported in \cite{crumbling}.  
    \item 
A second is to try to interpret the high-frequency power-law behaviour of the non-linear susceptibilities $\chi_3(\omega)$,  $\chi_5(\omega)$ measured in \cite{Alb16}, in the regime corresponding to the  ``excess wing'' for linear susceptibilities (see also \cite{Bau13}). If we follow the interpretation of \cite{scalliet_guiselin}, the excess wing is due to the early activation sites that then grow and propagate through facilitation. The corresponding behaviour of $\chi_3(\omega)$,  $\chi_5(\omega)$ could provide some information about the spatial structure of those ``activons''. Experimental measurements of (third harmonic) non-linear mechanical response would also be highly interesting \cite{fuchs, fuchs2}.   
    \item The third concerns rheology and fracture. As we pointed out in ref. \cite{RFOT2}, RFOT theory suggests a strong crossover from a high viscosity regime at low shear stress $\sigma$ to a low viscosity regime at higher shear stress, when the elastic energy $G_{\text{hf}} \sigma^2 \ell^3$ stored in a glassite exceeds the energy barrier $B$ given by Eq. \eqref{eq:Barrier}. Interestingly, the cross-over stress should {\it decrease} as temperature decreases \cite{RFOT2}. It is not clear what local theories would predict.\footnote{On this issue, see the extended discussion of P. Sollich, this volume.} Similarly, when a fracture propagates inside a super-cooled liquid close to the glass transition, one may also expect a brittle-ductile transition when concentrated stresses at the tip of the crack are able to ``liquify'' the glass ahead of the fracture front, with possibly hysteretic effects -- see Ref. \cite{babs}. The fracture surface left behind should correspondingly reveal an interesting crossover length between two roughness exponents \cite{physrep}. 
\end{itemize}

\subsection*{Acknowledgments} This piece is largely inspired by the recent review paper written in collaboration with Giulio Biroli \cite{CRAS}, to whom I owe much of my understanding of the glass transition. I also want to warmly thank L. Berthier, E. Bouchaud, F. Ladieu, S. Nagel, D.R. Reichman, C. Scalliet, P. Sollich, G. Tarjus, M. Wyart, H. Yoshino and F. Zamponi for many insightful discussions on all these issues over the years. Finally, I am grateful to the organizers of the Solvay conference, in particular M. Mézard and G. Parisi, for having invited me to such a memorable event. Finally, renewed thanks to L. Berthier, G. Biroli and P. Sollich for carefully reading my manuscript.


\begin{thebibliography}{99.}%
%
%

\bibitem{Nagel_Angell} Ediger, M. D., Angell, C. A., \& Nagel, S. R. (1996). Supercooled liquids and glasses. The journal of physical chemistry, 100(31), 13200-13212.

\bibitem{Angell} Angell, C. A. (1997). Entropy and fragility in supercooling liquids. Journal of research of the National Institute of Standards and Technology, 102(2), 171.


\bibitem{Xia} X. Xia, \& P. G. Wolynes, P. G., { Microscopic theory of heterogeneity and nonexponential relaxations in super-cooled liquids.} Physical Review Letters, 86(24), 5526 (2001)

\bibitem{AG} Adam, G., \& Gibbs, J. H. (1965). On the temperature dependence of cooperative relaxation properties in glass‐forming liquids. The journal of chemical physics, 43(1), 139-146.


\bibitem{KTW1} Kirkpatrick, T. R., \& Thirumalai, D. p-spin-interaction spin-glass models: Connections with the structural glass problem. Physical Review B, 36(10), 5388 (1987).

\bibitem{KTW2} Kirkpatrick, T. R., \& Thirumalai, D. Comparison between dynamical theories and metastable states in regular and glassy mean-field spin models with underlying first-order-like phase transitions. Physical Review A, 37(11), 4439 (1988).

\bibitem{KTW3} Kirkpatrick, T. R., Thirumalai, D., \& Wolynes, P. G. Scaling concepts for the dynamics of viscous liquids near an ideal glassy state. Physical Review A, 40(2), 1045 (1989).

\bibitem{biroli_mezard} G. Biroli and M. Mézard, (2001). Lattice Glass Models, Physical
Review Letters 88, 025501.

\bibitem{zamponi_book} Parisi, G., Urbani, P., \& Zamponi, F. (2020). Theory of simple glasses: exact solutions in infinite dimensions. Cambridge University Press.

\bibitem{CRAS} Biroli, G., \& Bouchaud, J. P. (2023). The RFOT Theory of Glasses: Recent Progress and Open Issues. Comptes Rendus. Physique, 24(S1), 1-15.


\bibitem{Montanari-Semerjian}
Montanari, A., \& Semerjian, G. (2006). {Rigorous inequalities between length and time scales in glassy systems}. Journal of statistical physics, 125(1), 23-54.


\bibitem{Bou04} J.-P. Bouchaud, and G. Biroli, {On the Adam-Gibbs-Kirkpatrick-Thirumalai-Wolynes scenario for the viscosity increase in glasses}, J. Chem. Phys. {\bf 121}, 7347 (2004).


\bibitem{RFOT1} Lubchenko, V., \& Wolynes, P. G. (2007). Theory of structural glasses and supercooled liquids. Annu. Rev. Phys. Chem., 58, 235-266.



\bibitem{wolynes-review} P.G. Wolynes and V. Lubchenko, (2012). {Structural glasses and super-cooled liquids: Theory, experiment, and applications}. John Wiley \& Sons.

\bibitem{RFOT2} G. Biroli, and J.-P. Bouchaud, in {Structural glasses and super-cooled liquids: theory, experiment, and applications},  P. G. Wolynes, V. Lubchenko, Eds. (Wiley, 2012), pp. 31-114.


\bibitem{Cam09} C. Cammarota,  A. Cavagna, G. Gradenigo, T. S. Grigera and P. Verrocchio, {Numerical determination of the exponents controlling the relationship between time, length, and temperature in glass-forming liquids}, J. Chem. Phys. {\bf 131}, 194901 (2009).


\bibitem{karmakar} Karmakar, S., Dasgupta, C., \& Sastry, S. (2014). Growing length scales and their relation to timescales in glass-forming liquids. Annu. Rev. Condens. Matter Phys., 5(1), 255-284.

\bibitem{pts1} L. Berthier, P. Charbonneau, D. Coslovich, A. Ninarello, M. Ozawa, S. Yaida, (2017). {Configurational entropy measurements in extremely super-cooled liquids that break the glass ceiling.} Proceedings of the National Academy of Sciences, 114(43), 11356-11361.


\bibitem{ozawa} Ozawa, M., Scalliet, C., Ninarello, A., \& Berthier, L. (2019). Does the Adam-Gibbs relation hold in simulated super-cooled liquids?. The Journal of chemical physics, 151(8), 084504.





\bibitem{KCM} D. Chandler, and J. P. Garrahan, {Dynamics on the Way to Forming Glass: Bubbles in Space-Time},
Annu. Rev. Phys. Chem. 61, 191-217 (2010).

\bibitem{bbreview} L. Berthier and G. Biroli, {Theoretical perspective on the glass transition and amorphous materials} Reviews of modern physics 83.2 (2011): 587.



\bibitem{Tarjus} Tarjus, G. (2011). An overview of the theories of the glass transition. Dynamical Heterogeneities in Glasses, Colloids, and Granular Media, 150, 39.



\bibitem{yoshino} Yoshino, H. (2012). Replica theory of the rigidity of structural glasses. The Journal of Chemical Physics, 136(21), 214108.

\bibitem{anderson} Anderson, P. W. (2018). Basic notions of condensed matter physics. CRC Press.


\bibitem{shoving}
J. C. Dyre, T. Christensen, N. B. Olsen. {Elastic models for the non-Arrhenius viscosity of glass-forming liquids}, Journal of non-crystalline solids, {\bf 352}(42-49) (2006). 



\bibitem{WC}
M. Wyart, M. Cates. {Does a growing static length scale control the glass transition?} Physical review letters, {\bf 119}(19), 195501 (2017).

\bibitem{Wnew} Ciamarra, M. P., Ji, W., \& Wyart, M. (2023). The energy cost of local rearrangements, not cooperative effects, makes glasses solid. arXiv preprint arXiv:2302.05150.

\bibitem{dyre_new} Dyre, J. C. (2024). Solid-that-flows picture of glass-forming liquids. The Journal of Physical Chemistry Letters, 15, 1603-1617.

\bibitem{Ritort_Sollich} Ritort, F., \& Sollich, P. Glassy dynamics of kinetically constrained models. Advances in physics, 52(4), 219-342 (2003).



\bibitem{Keys} Keys, A. S., Hedges, L. O., Garrahan, J. P., Glotzer, S. C., \& Chandler, D. (2011). Excitations are localized and relaxation is hierarchical in glass-forming liquids. Physical Review X, 1(2), 021013.

\bibitem{Wyart} Wyart, M. (2010). Correlations between vibrational entropy and dynamics in liquids. Physical review letters, 104(9), 095901.

\bibitem{franz}
Franz, S., \& Montanari, A. (2007). Analytic determination of dynamical and mosaic length scales in a Kac glass model. Journal of Physics A: Mathematical and Theoretical, 40(11), F251.


\bibitem{Tarjus_R} G. Tarjus, S. A. Kivelson, Z. Nussinov, and P. Viot, {The frustration-based approach of super-cooled liquids and the glass transition: a review and critical assessment}, J. Phys: Cond. Matt. {\bf 17}, R1143-R1182 (2005).

\bibitem{FP-potential} S. Franz and G. Parisi, {Recipes for metastable states in spin glasses}. Journal de Physique I 5.11 (1995): 1401-1415.

\bibitem{berthiernew} Berthier, L. (2021). Self-induced heterogeneity in deeply super-cooled liquids. Physical Review Letters, 127(8), 088002.

\bibitem{wolynes-rfim}
Stevenson, J. D., Walczak, A. M., Hall, R. W., \& Wolynes, P. G. (2008). Constructing explicit magnetic analogies for the dynamics of glass forming liquids. The Journal of chemical physics, 129(19), 194505.


\bibitem{franz-rfim} S. Franz, G. Parisi, F. Ricci-Tersenghi, T. Rizzo, (2011). { Field theory of fluctuations in glasses}. The European Physical Journal E, 34(9), 1-17.


\bibitem{RFIM1}
G. Biroli, C. Cammarota, G. Tarjus, and M. Tarzia, (2018). {Random-field Ising-like effective theory of the glass transition. I. Mean-field models.} Physical Review B, 98(17), 174205.

\bibitem{RFIM2}
G. Biroli, C. Cammarota, G. Tarjus, G. and M. Tarzia, (2018). {Random field Ising-like effective theory of the glass transition. II. Finite-dimensional models.} Physical Review B, 98(17), 174206.

\bibitem{guiselin} B. Guiselin, L. Berthier, \& G. Tarjus, (2022). {Statistical mechanics of coupled super-cooled liquids in finite dimensions}. SciPost Physics, 12(3), 091.

\bibitem{Cavagna} G. Biroli,  J. P. Bouchaud, A. Cavagna, T. S. Grigera,  \& P. Verrocchio. {Thermodynamic signature of growing amorphous order in glass-forming liquids.} Nature Physics, 4(10), 771-775 (2008).





\bibitem{dzero} M. Dzero, J. Schmalian, P.G. Wolynes, P. G. {Activated events in glasses: The structure of entropic droplets}. Physical Review B, {\bf 72}(10), 100201 (2005).



\bibitem{scalliet}
L. Berthier, M. Ozawa, C. Scalliet, {Configurational entropy of glass-forming liquids}. The Journal of chemical physics, {\bf 150}(16),  (2019) 160902.



\bibitem{Torchinsky} Torchinsky, D. H., Johnson, J. A., \& Nelson, K. A. (2009). A direct test of the correlation between elastic parameters and fragility of ten glass formers and their relationship to elastic models of the glass transition. The Journal of chemical physics, 130(6).


\bibitem{swap}
A. Ninarello, L. Berthier, D. Coslovich. {Models and algorithms for the next generation of glass transition studies} Physical Review X, {\bf 7}(2), 021039 (2017).

\bibitem{Gava} Gavazzoni, C., Brito, C., \& Wyart, M. (2023). Testing theories of the glass transition with the same liquid, but many kinetic rules. arXiv preprint arXiv:2308.00196.


\bibitem{crumbling} L. Berthier, G. Biroli, J.-P. Bouchaud, G. Tarjus,  { Can the glass transition be explained without a growing static length scale?} The Journal of chemical physics, {\bf 150}(9), 094501 (2019). 


\bibitem{goldstein} Goldstein, M. (1969). Viscous liquids and the glass transition: a potential energy barrier picture. The Journal of Chemical Physics, 51(9), 3728-3739.


\bibitem{coslovich} Coslovich, D., Ninarello, A., \& Berthier, L. (2019). A localization transition underlies the mode-coupling crossover of glasses. SciPost Physics, 7(6), 077.


\bibitem{BBT} Biroli, G., Bouchaud, J. P., \& Tarjus, G. (2005). Are defect models consistent with the entropy and specific heat of glass formers?. The Journal of chemical physics, 123(4).

\bibitem{brito} Brito, C., Lerner, E., \& Wyart, M. (2018). Theory for swap acceleration near the glass and jamming transitions for continuously polydisperse particles. Physical Review X, 8(3), 031050.

\bibitem{ikeda} Ikeda, H., Zamponi, F., \& Ikeda, A. (2017). Mean field theory of the swap Monte Carlo algorithm. The Journal of chemical physics, 147(23), 234506.



\bibitem{szamel} Szamel, G. (2018). Theory for the dynamics of glassy mixtures with particle size swaps. Physical Review E, 98(5), 050601.


\bibitem{HL} Hébraud, P., \& Lequeux, F. (1998). Mode-coupling theory for the pasty rheology of soft glassy materials. Physical review letters, 81(14), 2934.



\bibitem{DH1} Berthier, L., Biroli, G., Bouchaud, J. P., \& Jack, R. L. (2011). Overview of different characterisations of dynamic heterogeneity. Dynamical heterogeneities in glasses, colloids, and granular media, 150, 68.

\bibitem{candelier} Candelier, R., Widmer-Cooper, A., Kummerfeld, J. K., Dauchot, O., Biroli, G., Harrowell, P., \& Reichman, D. R. (2010). Spatiotemporal hierarchy of relaxation events, dynamical heterogeneities, and structural reorganization in a supercooled liquid. Physical review letters, 105(13), 135702.




\bibitem{chacko}
R. Chacko, F. Landes, G. Biroli, O. Dauchot, A. Liu, and D. R. Reichman (2021). {Elastoplasticity Mediates Dynamical Heterogeneity Below the Mode-Coupling Temperature.} Physical Review Letters, 127(4), 048002.


\bibitem{scalliet_guiselin} Scalliet, C., Guiselin, B., \& Berthier, L. (2022). Thirty milliseconds in the life of a supercooled liquid. Physical Review X, 12(4), 041028.


\bibitem{Tahaei} Tahaei, A., Biroli, G., Ozawa, M., Popović, M., \& Wyart, M. (2023). Scaling Description of Dynamical Heterogeneity and Avalanches of Relaxation in Glass-Forming Liquids. Physical Review X, 13(3), 031034.


\bibitem{excess_wings} Guiselin, B., Scalliet, C., \& Berthier, L. (2022). Microscopic origin of excess wings in relaxation spectra of supercooled liquids. Nature Physics, 18(4), 468-472.


\bibitem{WW} Wisitsorasak, A., \& Wolynes, P. G. (2014). Dynamical heterogeneity of the glassy state. The Journal of Physical Chemistry B, 118(28), 7835-7847.


\bibitem{berthier2023} Herrero, C., \& Berthier, L. (2023). Direct numerical analysis of dynamic facilitation in glass-forming liquids. arXiv preprint arXiv:2310.16935.



\bibitem{IMCT} Biroli, G., Bouchaud, J. P., Miyazaki, K., \& Reichman, D. R. (2006). Inhomogeneous mode-coupling theory and growing dynamic length in supercooled liquids. Physical review letters, 97(19), 195701.

\bibitem{DH2} Berthier, L., Biroli, G., Bouchaud, J. P., Kob, W., Miyazaki, K., \& Reichman, D. R. (2007). Spontaneous and induced dynamic fluctuations in glass formers. I. General results and dependence on ensemble and dynamics. The Journal of chemical physics, 126(18).



\bibitem{nishikawa} Nishikawa, Y., \& Berthier, L. (2023). Collective relaxation dynamics in a three-dimensional lattice glass model. arXiv preprint arXiv:2307.08110.


\bibitem{Alb16} S. Albert, Th. Bauer, M. Michl, G. Biroli, J.-P. Bouchaud, A. Loidl, P. Lunkenheimer, R. Tourbot, C. Wiertel-Gasquet, and F. Ladieu, {Fifth-order susceptibility unveils growth of thermodynamic amorphous order in glass-formers}, Science {\bf 352}, 1308 (2016).


\bibitem{pinning-expt} Gokhale, S., Hima Nagamanasa, K., Ganapathy, R., \& Sood, A. K. (2014). Growing dynamical facilitation on approaching the random pinning colloidal glass transition. Nature communications, 5(1), 1-7.

\bibitem{Das0} Das, R., Chakrabarty, S., \& Karmakar, S. (2017). Pinning susceptibility: a novel method to study growth of amorphous order in glass-forming liquids. Soft matter, 13(38), 6929-6937.

\bibitem{Das} Das, R., Bhowmik, B. P., Puthirath, A. B., Narayanan, T. N., \& Karmakar, S. (2021). Soft-Pinning: Experimental Validation of Static Correlations in super-cooled Molecular Glass-forming Liquids. arXiv preprint arXiv:2106.06325.

\bibitem{Bou05} J.-P. Bouchaud, and G. Biroli, {Nonlinear susceptibility in glassy systems: A probe for cooperative dynamical length scales}, Phys. Rev. B {\bf 72}, 064204 (2005).

\bibitem{BBL} Biroli, G., Bouchaud, J. P., \& Ladieu, F. (2021). Amorphous Order and Nonlinear Susceptibilities in Glassy Materials. The Journal of Physical Chemistry B, 125(28), 7578-7586.


\bibitem{LadieuNew} Bertin, E., \& Ladieu, F. (2023). Nonlinear dielectric response in glasses: restoring forces and avoided spin-glass criticality. arXiv preprint arXiv:2312.04267.


\bibitem{Bru12} C. Brun, F. Ladieu, D. L'H\^{o}te, G. Biroli, and J.-P. Bouchaud, {Evidence of growing spatial correlations during the aging of glassy glycerol}, Phys. Rev. Lett. {\bf 109}, 175702 (2012).


\bibitem{guiselin_overlaps} Guiselin, B., Tarjus, G., \& Berthier, L. (2022). Static self-induced heterogeneity in glass-forming liquids: Overlap as a microscope. The Journal of Chemical Physics, 156(19), 194503.

\bibitem{wolynes_2013} Rabochiy, P., Wolynes, P. G., \& Lubchenko, V. (2013). Microscopically based calculations of the free energy barrier and dynamic length scale in supercooled liquids: The comparative role of configurational entropy and elasticity. The Journal of Physical Chemistry B, 117(48), 15204-15219.

\bibitem{cooling} Chacko, R. N., Landes, F. P., Biroli, G., Dauchot, O., Liu, A. J., \& Reichman, D. R. (2023). Dynamical Facilitation Governs the Equilibration Dynamics of Glasses. arXiv preprint arXiv:2312.15069.
 

\bibitem{Bau13} Th. Bauer, P. Lunkenheimer, S. Kastner, and A. Loidl, {Nonlinear dielectric response at the excess wing of glass-forming liquids}, Phys. Rev. Lett. {\bf 110}, 107603 (2013).


\bibitem{fuchs}
C\'ardenas, H., Frahsa, F., Fritschi, S., Nicolas, A., Papenkort, S., Voigtmann, T., \& Fuchs, M. (2017). Nonlinear mechanical response of super-cooled melts under applied forces. The European Physical Journal Special Topics, 226(14), 3039-3060.

\bibitem{fuchs2}
Seyboldt, R., Merger, D., Coupette, F., Siebenbürger, M., Ballauff, M., Wilhelm, M., \& Fuchs, M. (2016). Divergence of the third harmonic stress response to oscillatory strain approaching the glass transition. Soft Matter, 12(43), 8825-8832.

\bibitem{babs} Gimenes, G. E., \& Bouchaud, E. (2018). Flow and fracture near the sol–gel transition of silica nanoparticle suspensions. Soft Matter, 14(39), 8036-8043.

\bibitem{physrep} Bonamy, D., \& Bouchaud, E. (2011). Failure of heterogeneous materials: A dynamic phase transition?. Physics Reports, 498(1), 1-44.


\end{thebibliography}
\end{document}